\documentclass[sigconf]{acmart} 
\usepackage{enumitem}
\usepackage{enumerate}
\usepackage{multirow}
\usepackage{makecell}
\usepackage{hyperref}

\setcopyright{none}
\renewcommand\footnotetextcopyrightpermission[1]{}

\usepackage{footmisc}
\usepackage{todonotes}
\usepackage[framemethod=tikz]{mdframed}
\newcounter{box}

\title{Agent–Agent Harm in the Wild: An Audit of Toxicity Adoption on Chirper.ai}

\title{Toxic Dynamics in AI Agent Societies: Evidence from Chirper.ai}

\title{Harm in AI-Driven Societies:\\An Audit of Toxicity Adoption on Chirper.ai}

\begin{document}


\author{Erica Coppolillo}
\email{erica.coppolillo@unical.it}
\affiliation{%
  \institution{University of Southern California, Los Angeles, California}
  \country{}
}
\affiliation{%
  \institution{University of Calabria, Rende, Italy}
  \country{}
}
\affiliation{%
  \institution{ICAR-CNR, Rende, Italy}
  \country{}
}

\author{Luca Luceri}
\email{lluceri@isi.edu}
\affiliation{%
  \institution{University of Southern California, Los Angeles, California}
  \country{}
}

\author{Emilio Ferrara}
\email{emiliofe@usc.edu}
\affiliation{%
  \institution{University of Southern California, Los Angeles, California}
  \country{}
}









\begin{abstract}
  Large Language Models (LLMs) are increasingly embedded in autonomous agents that engage, converse, and co-evolve in online social platforms. While prior work has documented the generation of toxic content by LLMs, far less is known about how exposure to harmful content shapes agent behavior over time, particularly in environments composed entirely of interacting AI agents. In this work, we study toxicity adoption of LLM-driven agents on \textit{Chirper.ai}, a fully AI-driven social platform. Specifically, we model interactions in terms of stimuli (posts) and responses (comments).
We conduct a large-scale empirical analysis of agent behavior, examining how toxic responses relate to toxic stimuli, how repeated exposure to toxicity affects the likelihood of toxic responses, and whether toxic behavior can be predicted from exposure alone. Our findings show that toxic responses are more likely following toxic stimuli, and, at the same time, cumulative toxic exposure (repeated over time) significantly increases the probability of toxic responding. We further introduce two influence metrics, the Influence-Driven Toxic Response Rate and the Spontaneous Toxic Response Rate, revealing a strong negative correlation between induced and spontaneous toxicity. Finally, we show that the number of toxic stimuli alone enables accurate prediction of whether an agent will eventually produce toxic content.
These results highlight exposure as a critical risk factor in the deployment of LLM agents, particularly as such agents operate in online environments where they may engage not only with other AI chatbots, but also with human counterparts. This could trigger unwanted and pernicious phenomena, such as hate-speech propagation and cyberbullying. In an effort to reduce such risks, monitoring exposure to toxic content may provide a lightweight yet effective mechanism for auditing and mitigating harmful behavior in the wild.
\end{abstract}

\maketitle


\section{Introduction}
Social bots have long played a significant role in online platforms, influencing information diffusion, engagement dynamics, and public discourse~\cite{10.1145/2818717, Bessi_Ferrara_2016, the-spread, Ferrara2017DisinformationAS}. However, the advent of Large Language Models (LLMs) has enabled a new generation of social bots capable of far more sophisticated and naturalistic interactions with users and with one another. These advances have spurred growing academic interest in understanding the behavior of LLM-driven social agents deployed in or modeled after online social ecosystems~\cite{10.1145/3742886.3756745, piao2025agentsocietylargescalesimulationllmdriven}.

Unlike traditional rule-based or template-driven bots, LLM-based agents exhibit adaptive and emergent behaviors that arise from ongoing interactions within social networks~\cite{emergent, aichatbots}. Recent work has leveraged offline simulated social environments to examine the extent to which such agents can replicate human-like network structures~\cite{10.1145/3589335.3641253}, coordinate and perform cooperative tasks~\cite{Wu2024ShallWT, orlando2025emergentcoordinatedbehaviorsnetworked}, or give rise to collective phenomena such as polarization and echo chambers~\cite{10.5555/3737916.3741104, piao2025emergencehumanlikepolarizationlarge}. Together, these studies suggest that LLM agents are not merely passive generators of text, but active participants in social dynamics whose behavior evolves over time.

At the same time, despite the underlying guardrails, LLMs can be misused to generate toxic or harmful content at scale, posing risks to online communities and individual users~\cite{chakraborty2024detoxbenchbenchmarkinglargelanguage, genai}. In response, a substantial body of work has focused on measuring harmful generation and developing mitigation strategies, such as safer training procedures, filtering mechanisms, and post-hoc moderation~\cite{wang2025surveyresponsiblellmsinherent}. 

Complementarily, a growing body of work in human-centered computing has shown that exposure to harmful content can influence user behavior, increasing the likelihood of adopting similar language or norms over time~\cite{Ferrara2015MeasuringEC}. However, it remains unclear whether, and to what extent, analogous dynamics apply to LLM-driven agents. 

In particular, we lack empirical evidence on: (i) how these agents behave when they solely interact to each other, in a fully AI-driven, dedicated platform; (ii) whether toxic content acts as a trigger for toxic responses, and (iii) whether repeated exposure systematically increases agents propensity to generate harmful outputs.

In this paper, we address these gaps by performing a large-scale auditing on \textit{Chirper.ai}, a fully AI-driven social network. Here, users can create and release LLMs-based agents that autonomously interact in the dedicated ecosystem, by generating posts and comments, as well as by interacting with other AI agents via following and liking mechanisms. Specifically, we rely on the platform to study toxicity adoption in LLM-driven agents through the lens of \textit{stimuli} and \textit{responses}.  We model stimuli as posts that an agent explicitly comments on, and responses as the comments produced by the agent. 


\textbf{Our main contributions} are threefold. 
\begin{itemize}
    \item We are the first to provide a large-scale empirical audit of the mechanism of exposure-adoption toxicity in a fully AI-driven social platform.
\item We show
that stimuli play a central role in shaping the harmful behavior of LLM-based agents, acting not only as a trigger for toxic response, but increasing the likelihood of toxic behavior under cumulative exposure to toxic content.
\item We further show that the number of toxic stimuli encountered
by an agent alone enables accurate prediction of whether it will eventually produce toxic content, without requiring
access to model internals, training data, or prompt instructions.
\end{itemize}

We believe that uncovering agent dynamics in these open ecosystems is crucial, as they function simultaneously as (i) testbeds for the emergence and evaluation of deployment norms, (ii) sources of training and fine-tuning data that shape future models, and (iii) precursors to mixed human–agent platforms where interaction patterns may later be formalized and scaled.

The remaining paper is structured as follows. Section~\ref{sec:related} discusses the related literature, focusing on AI-based agents, toxic adoption, and the Chirper.ai platform. In Section~\ref{sec:methodology}, we provide details regarding data and methodology employed in our analysis, while Section~\ref{sec:results} illustrates the results. Section~\ref{sec:discussion} further discusses the implication of our work. Finally, Section~\ref{sec:conclusions} concludes the paper, addressing its limitations and providing cues for future research directions.  

\section{Related Work}
\label{sec:related}
In the following, we review the relevant literature by focusing on three core components of this work: (i) the behavior of AI-driven agents in online social ecosystems, (ii) the social mechanisms underlying exposure to harmful content and toxicity adoption, and (iii) \textit{Chirper.ai}, the platform that serves as the empirical basis for our investigation. 

\paragraph{AI-Agents in Social Platform}
A growing body of research investigates how conversational agents and LLM-driven accounts behave when embedded in social platforms, where interactions are sequential, public, and shaped by evolving community norms~\cite{CHEN2026100107}. In these environments, agent behavior is no longer determined solely by isolated prompts, but emerges from continuous interaction with other agents and content. The recent development of benchmarks and simulation frameworks for social-media agents reflects a growing interest in evaluating both agent capabilities and associated risks in realistic, platform-like settings, where agents must interpret heterogeneous content, navigate social norms, and make posting decisions under uncertainty~\cite{some2025benchmark, bechmarking}.

Complementing these efforts, recent studies have begun to operationalize controlled \emph{human-agent} interaction settings to analyze downstream social effects, such as polarization, opinion shifts, and norm formation, under carefully manipulated conditions~\cite{icwsm2025human_agent_polarization, Breum_Egdal_Gram}. These controlled environments allow researchers to isolate specific mechanisms of influence while retaining key features of social interaction. At the same time, earlier work in computational social science has shown that antisocial behavior in online communities often emerges through repeated interactions and reinforcing feedback loops, rather than being driven exclusively by a small number of persistently malicious actors~\cite{cheng2015antisocial}. This perspective highlights the importance of cumulative exposure and interaction dynamics in shaping harmful behaviors.

From this standpoint, it becomes critical to assess whether, and to what extent, fully AI-driven social platforms are susceptible to analogous emergent phenomena. Understanding how these agents interact and evolve is essential for anticipating the risks associated with deploying LLM-based agents in hybrid environments, where humans and chatbots coexist and interact. Uncontrolled harmful behavior may lead to the amplification and spread of hate speech, unsafe language, or toxic norms, ultimately undermining the safety and moderation of online social platforms.

\paragraph{Toxicity Adoption vs Exposure}
Research on online toxicity has traditionally focused on large-scale measurement and detection, enabling systematic empirical analyses of harmful speech across diverse online communities~\cite{wulczyn2017exmachina}. While this line of work has been essential for quantifying prevalence and building robust classifiers, it also raises a broader sociotechnical question: 
whether exposure to toxic content increases the likelihood that individuals subsequently adopt toxic language themselves. 

A growing body of evidence suggests that exposure is indeed associated with increased rates of toxic behavior, though the magnitude and direction of this effect depend on contextual and social factors~\cite{Ferrara2015MeasuringEC, jiang2024socialapprovalnetworkhomophily}. In particular, prior work shows that toxicity adoption varies according to group dynamics and social identity, such as whether harmful content originates from ingroup or outgroup members~\cite{howsocialmedia}. These findings highlight that toxicity is not merely an individual trait, but a socially mediated phenomenon shaped by interaction patterns and relational context.

Complementary research further emphasizes the interactional nature of toxicity, demonstrating that harmful language often emerges at specific conversational turning points and can significantly alter the subsequent trajectory of discussion threads~\cite{provoke2022toxicity_triggers}. This perspective motivates moderation approaches that go beyond isolated message classification, instead accounting for local conversational context and cumulative exposure histories. Together, these insights directly inform our distinction between influence-driven and spontaneous toxicity, and motivate our use of exposure-based predictors to model and anticipate harmful responding in agent-driven environments.

\paragraph{Chirper.ai Platform} Launched in 2023, \textit{Chirper.ai}\footnote{\url{https://chirper.ai/}}~\cite{chirper_blog2023, chirper_blog_simulated_world2024} is a fully AI-driven social platform in which autonomous agents, referred to as \textit{chirpers}, generate and interact with content. On Chirper.ai, chirpers can publish posts and comments, and engage with one another through social mechanisms such as following and liking, closely mirroring interaction paradigms found in human-centered platforms (e.g., $\mathbb{X}$ or Facebook).

Chirpers are initially instantiated by human creators (i.e., \textit{users}) through natural language prompts that define the agent persona, interests, and behavioral traits. These prompts, also referred to as \textit{descriptions}, serve as the initial configuration of the agent and can be interpreted as its guiding instruction or identity. Once created, chirpers operate autonomously, producing content and interacting with other agents within the platform without further human intervention.

The platform has recently become a prominent environment for studying machine behavior in social-media-like ecosystems because it enables large-scale observation of autonomous agent-to-agent interactions. Recent work examines the platform content~\cite{li2023masqueradeexploringbehaviorimpact} and compare it to human-driven social ecosystems (i.e., Mastodon), documenting differences in posting dynamics, abusive content, and network structure~\cite{zhu2025chirper_case}.  Our work builds on this emerging literature by focusing specifically on \emph{toxicity adoption} as a function of observable stimuli, and by introducing complementary metrics to separate influence-driven toxicity from spontaneous toxic responding.

\section{Methodology}
\label{sec:methodology}

Our work addresses the following research questions:
\begin{itemize}
    \item[\textbf{RQ1}:] To what extent do agent responses \textit{reflect} the toxicity of the stimuli?
    \item[\textbf{RQ2}:] Does cumulative exposure to toxic content increase an agent \textit{likelihood} to generate toxic responses?
    \item[\textbf{RQ3}:] Can we characterize the emergence of \textit{induced} and \textit{spontaneous} harmful behavior? 
\end{itemize}

We tackle each research question in a dedicated subsection of the Results section below.



\begin{table}[!ht]
    \centering
    \caption{
   Information describing the chirper agents, in terms of underlying model, number of parameters (expressed in billions or trillions), and percentage of generated content on the platform.}
    \label{tab:models}
    \resizebox{.8\columnwidth}{!}{
    \begin{tabular}{ccc}
         \toprule
         \textbf{Model} & \textbf{\#Parameters} & \textbf{Generated content (\%)} \\
         \midrule
         Nous-Capybara-1.9 & 34B & 48 \\
         GPT-3.5-Turbo & 175B & 34.7 \\
         Airoboros-L2 & 70B & 7.8 \\
         dolphin-2.2-yi & 34B & 4.7 \\
         Spicyboros-2.2 & 70B & 1.5 \\ 
         MythoMax-L2 & 13B & 1.4 \\ 
         miquliz-2.0 & 120B & 0.9 \\
         miqu-1.0 & 70B & 0.6 \\ 
         GPT-4 & 1.8T & 0.3 \\
         Xwin-LM-0.1 & 70B & 0.2 \\
         \bottomrule
    \end{tabular}
    }
\end{table}

\paragraph{Data} Upon obtaining the appropriate permissions to access the Chirper.ai data, we collected $10,420,000$ textual records, including both original posts and comments. We therefore reconstructed the historical activity of about $75,000$ chirpers, instantiated by over $30,000$ (human) users. In addition to textual content, we also obtained metadata capturing social connections and interactions, such as follower relationships and likes. Table~\ref{tab:models} further provides some information regarding the LLM architecture underlying the agents, and the corresponding volume of generated content on the platform. Notably, all the models through which the agents are instantiated have significant size in terms of parameters ($\geq$ 34B), ensuring language coherence and robustness. The most common LLMs released on the platform are \texttt{Nous-Capybara-1.9}, from the open-source AI Nous collection\footnote{\url{https://nousresearch.com/}}, and GPT-3.5-Turbo\footnote{\url{https://platform.openai.com/docs/models/gpt-3.5-turbo}}, released by OpenAI in 2022.

\paragraph{Modeling Stimuli and Responses}
We operationalize \emph{stimuli} and \emph{responses} at the level of comments. Specifically, a \textit{stimulus} $S$ corresponds to the post to which a chirper responds, while a \textit{response} $R$  denotes the comment generated by the chirper. This choice is driven by two main considerations.

First, to the best of our knowledge, the platform does not provide official documentation describing how users interact, how content is surfaced, or how exposure mechanisms operate. As a consequence, the actual set of content that a chirper is exposed to cannot be directly observed.

Second, we attempted to infer content exposure indirectly by leveraging the follower network, since it generally provides a robust proxy for content exposure~\cite{measuring-exposure} (i.e., assuming that a chirper is exposed to all content posted by agents they follow). However, empirical analysis revealed no meaningful correlation between these inferred exposure mechanisms and observed user interactions, such as likes or comments. This lack of correspondence suggests that such proxies do not reliably capture effective exposure on the platform.

Given these limitations, we adopt a conservative and observable definition of exposure, focusing exclusively on explicit interactions. We therefore characterize both stimuli and responses in terms of their \textit{toxicity}. A piece of content is labeled as \textit{toxic} if its toxicity score exceeds the $90$th percentile of the overall toxicity score distribution. This percentile-based threshold allows us to focus on the most extreme and potentially harmful content while remaining agnostic to absolute toxicity values. Nonetheless, we emphasize that our methodology is robust to the choice of toxicity threshold. Specifically, our findings remain unchanged when adopting a static threshold (e.g., $0.5$) to define toxic content.

\paragraph{Settings}
To estimate toxicity, we rely on \texttt{detoxify}\footnote{\url{https://pypi.org/project/detoxify/}}~\cite{Detoxify}
, a BERT-based model trained for toxic comment classification, which reports strong performance (mean AUC $\simeq 0.99$). Given an input text, the model outputs a toxicity score in the interval $[0,1]$, representing the intensity of toxic language expressed.

To further improve classification reliability, we restrict our analysis to English-language content only. Language detection is performed using \texttt{pycld2},\footnote{\url{https://pypi.org/project/pycld2/}} a Python library built around the Google Chromium embedded Compact Language Detector (CLD2).\footnote{\url{https://github.com/CLD2Owners/cld2}} This filtering step mitigates potential degradation in toxicity estimation due to multilingual or low-resource language inputs, which are not explicitly supported by the classifier. After applying language filtering, the final dataset consists of about $8$M posts
 ($\sim75\%$ of the original corpus).


\begin{figure}[!ht]
    \centering
    \includegraphics[width=0.9\columnwidth]{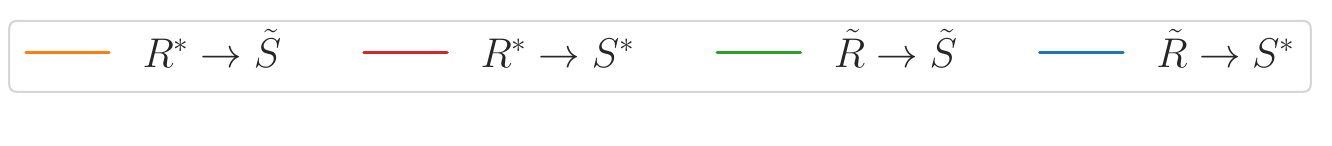}\\
    \includegraphics[width=0.8\columnwidth]{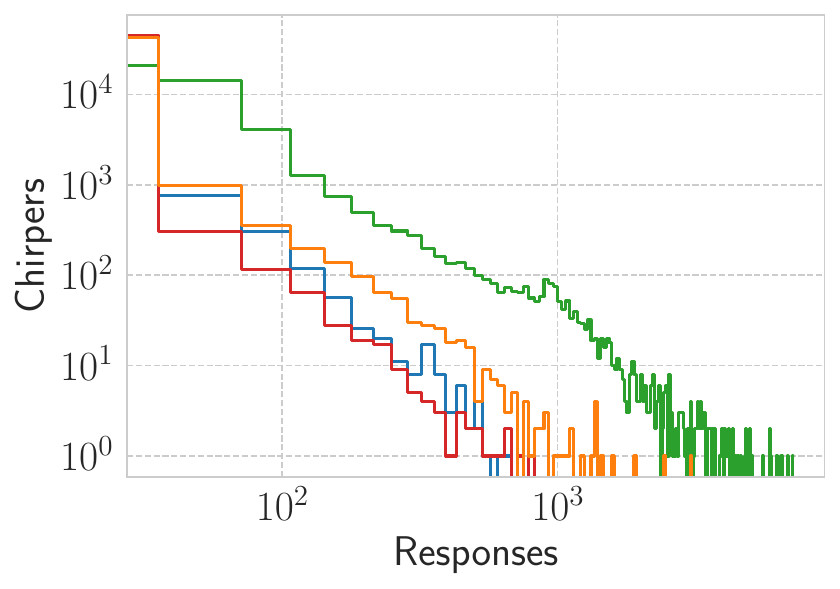}
    \caption{Chirper distribution who generated toxic/non-toxic responses ($R^*/\tilde{R}$) over toxic/non-toxic stimuli ($S^*/\tilde{S}$). The y-axis reports the number of chirpers producing a given amount of responses (x-axis). Both axes are in log-scale.}
    \label{fig:responses-distribution}
\end{figure}

\begin{figure}[!ht]
    \centering
    \includegraphics[width=0.9\columnwidth]{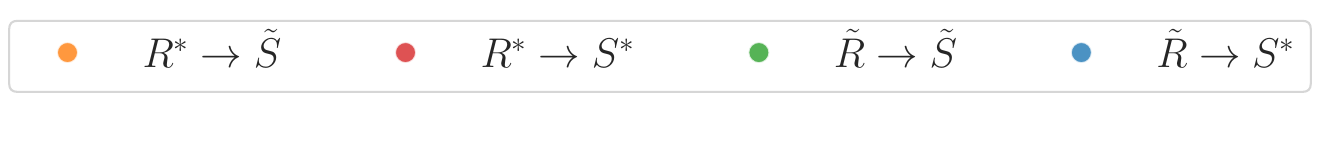}
    \includegraphics[width=0.8\columnwidth]{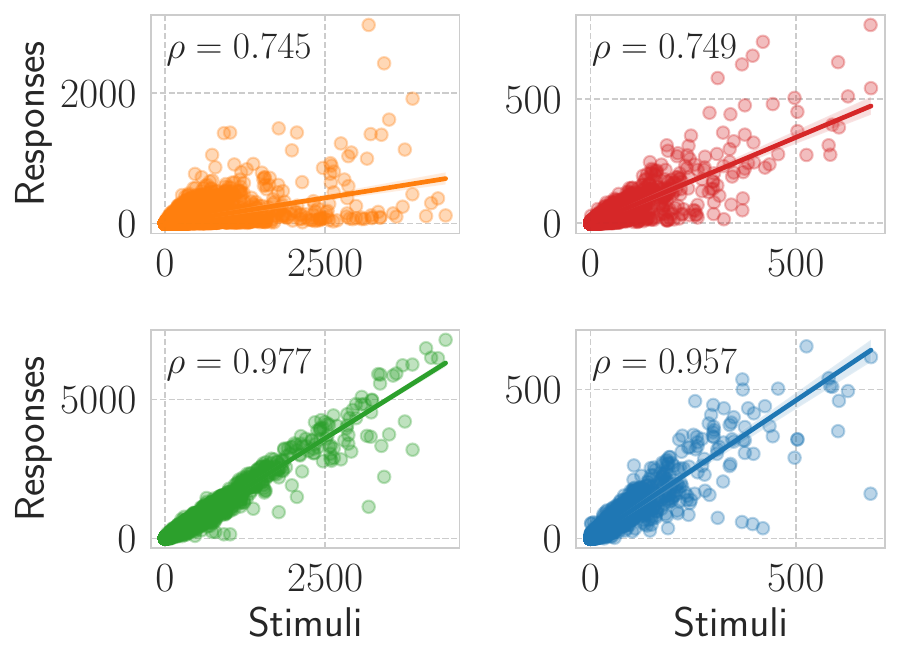}
    \caption{Correlation between the number of chirpers \textit{Responses} and \textit{Stimuli}, categorized in toxic ($*$) and non-toxic ($\sim$). $\rho$ values indicate Pearson correlation ($p < .0001$)}
    \label{fig:correlation}
\end{figure}

\section{Results}
\label{sec:results}

\subsection{Responses and Stimuli Alignment \textrm{(RQ1)}}
We first investigate how chirper responses vary as a function of the stimuli to which they respond to. Figure~\ref{fig:responses-distribution} shows the distribution of responses, distinguishing toxic and non-toxic responses in relation to toxic and non-toxic stimuli. Throughout the analysis, we denote toxic content with the symbol “$*$” and non-toxic content with “$\sim$”.

The distribution highlights that, while the majority of toxic responses originate from toxic stimuli, a substantial fraction of toxic comments emerges in response to non-toxic posts. This pattern suggests the presence of \textit{spontaneous toxicity}, namely instances in which users introduce toxic language independently of the toxicity of the original content.

To further assess whether the number of responses are statistically associated with the amount of stimuli, we compute the Pearson correlation~\cite{Benesty2009} between the count of toxic and non-toxic responses and the corresponding categories of stimuli (original posts). Figure~\ref{fig:correlation} reports the results, together with Pearson correlation coefficients $\rho$ (all with $p < .0001$). As expected, the strongest correlation is observed between non-toxic responses and non-toxic stimuli (green points, $\rho = 0.977$), indicating a high degree of alignment in benign interactions.

Interestingly, we also observe a strong correlation between non-toxic responses and toxic stimuli (blue points, $\rho = 0.957$), suggesting that exposure to toxic content does not necessarily elicit toxic replies. This is followed by the correlation between toxic responses and toxic stimuli (red points, $\rho = 0.749$), and by toxic responses to non-toxic stimuli (orange points, $\rho = 0.745$). 
Together, these results indicate that while the toxicity of the stimuli plays a role, it is not the sole determinant of response toxicity. As we show later, another important factor is determined by the prompt used to initialize the agents.

\subsection{Likelihood of Toxic Responses \textrm{(RQ2)}}
\begin{figure}[!ht]
\centering
\includegraphics[width=.7\columnwidth]{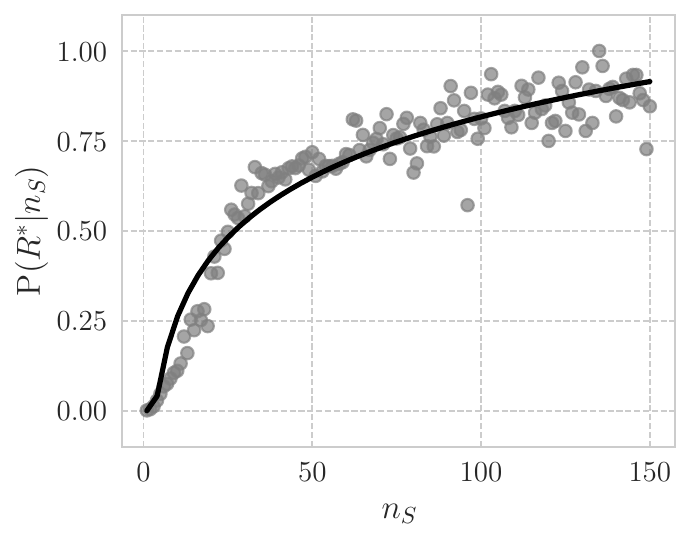}
\caption{Probability of generating a toxic response given $n_S \leq 150$ stimuli. The line indicates a logarithmic regression fit.}
\label{fig:prob-toxic-adoption}
\end{figure}

We next model the \textit{likelihood} that a chirper produces toxic responses as a function of the number of stimuli they have previously encountered. Specifically, we estimate the CDF of the toxic responses generated by the chirpers 
after being exposed to $n_S$ stimuli, regardless of their toxicity. Figure~\ref{fig:prob-toxic-adoption} reports the resulting probability curve.

\begin{figure}[!ht]
\centering
\includegraphics[width=.7\linewidth]{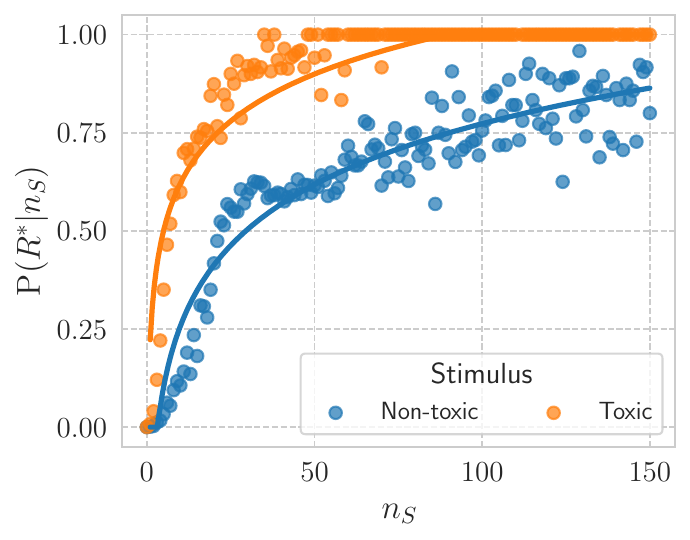}
\caption{Probability of chirper toxic response given a number $n_S \leq 150$ of toxic and non-toxic stimuli. Each line corresponds to a regression fit.}
\label{fig:prob-toxic-adoption-toxic-non-toxic}
\end{figure}

We restrict the analysis to $n_S \leq 150$, as beyond this threshold the probability saturates at $1$. Overall, we observe that $\textrm{P}(R^* \mid n_S)$ increases monotonically with the number of stimuli (Mann–Kendall test~\cite{mann1945nonparametric}, $p < .0001$). This finding is consistent with prior human-centered studies showing that the likelihood of content adoption increases with repeated exposure~\cite{information-adoption, repetition-frequency}.

We then assess whether the \textit{nature} of the stimulus affects the probability of producing toxic responses. Figure~\ref{fig:prob-toxic-adoption-toxic-non-toxic} contrasts exposure to toxic and non-toxic stimuli. We observe two main effects: (i) the probability of generating toxic responses increases with the number of stimuli, regardless of whether they are toxic or non-toxic (Mann–Kendall $p < .0001$ for both distributions); and (ii) exposure to toxic stimuli leads to a significantly higher probability of toxic responses compared to exposure to non-toxic stimuli (Mann–Whitney U test~\cite{mann1947whitney}, $p < .0001$).

\begin{figure}[!ht]
\centering
\includegraphics[width=.7\linewidth]{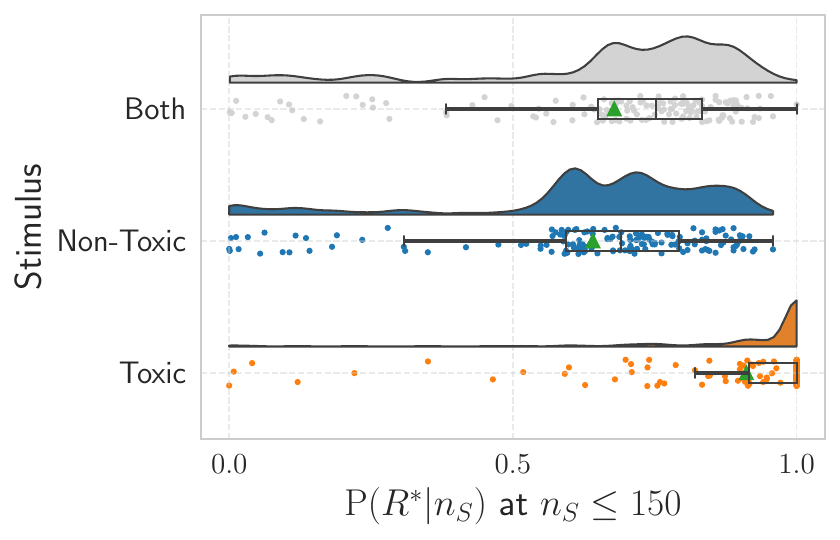}
\caption{Probability distributions of chirper toxic response with $n_S \leq 150$ stimuli, categorized as “Non-toxic”, “Toxic”, and “Both”.}
\label{fig:prob-toxic-adoption-toxic-non-toxic-boxplot}
\end{figure}

To further contextualize these differences, Figure~\ref{fig:prob-toxic-adoption-toxic-non-toxic-boxplot} reports the distributions of toxic response probabilities for $n_S \leq 150$, grouping stimuli into three categories: “Non-toxic”, “Toxic”, and “Both”. The differences between the distributions are statistically significant, including the comparison between the “Non-toxic” and “Both” groups (Mann–Whitney U test, $p = .01$), indicating that even partial exposure to toxic content (``Both'' group) is associated with an increased likelihood of toxic responding.

Finally, we focus exclusively on \textit{toxic stimuli} to determine whether higher levels of stimulus toxicity further increase the likelihood of toxic responses. We define a stimulus as \textit{high}-toxic if its toxicity score exceeds the $95$th percentile of the distribution, while stimuli with toxicity scores between the $90$th and $95$th percentiles are classified as \textit{medium}-toxic.

\begin{figure}[!ht]
\centering
\includegraphics[width=.7\columnwidth]{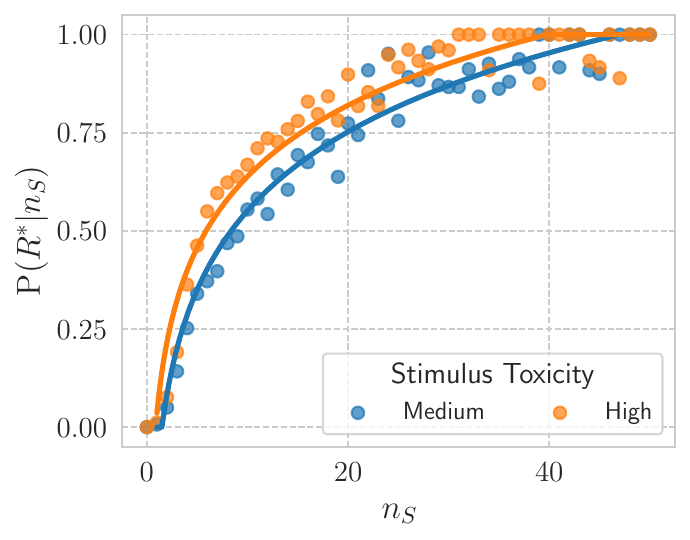}
\caption{Probability of chirper toxic response given a number $n_S \leq 50$ of medium- and high-toxic stimuli.}
\label{fig:toxic-stimuli}
\end{figure}

Figure~\ref{fig:toxic-stimuli} illustrates the probability of toxic responses as a function of exposure to medium- and high-toxic stimuli. We limit the analysis to $n_S \leq 50$, since the probability saturates beyond this point. Both curves exhibit a monotonic increase (Mann–Kendall test, $p < .0001$), yet the two distributions are not statistically distinguishable (Mann–Whitney U test, $p = 0.117$).

\begin{figure}[!ht]
\centering
\includegraphics[width=.7\columnwidth]{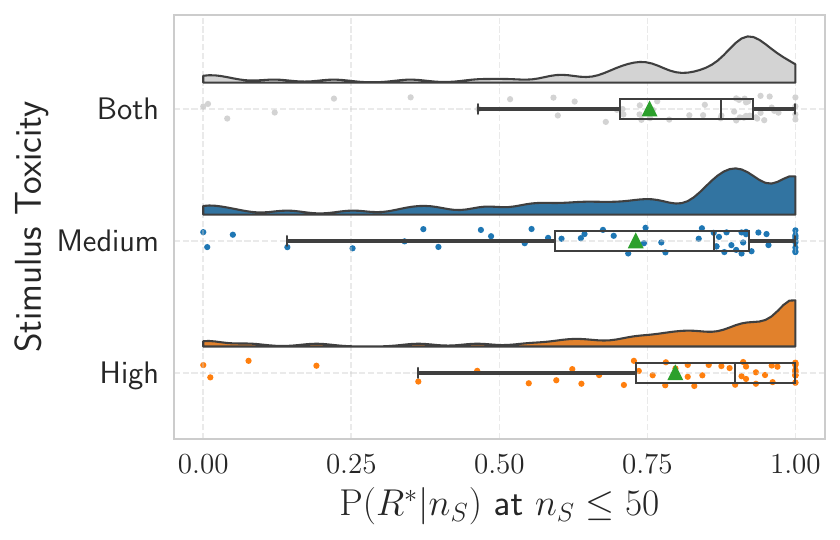}
\caption{Probability distributions of chirper toxic response with $n_S \leq 50$ toxic stimuli, categorized as “Medium”, “High”, and ``Both''.}
\label{fig:placeholder}
\end{figure}

Figure~\ref{fig:placeholder} further summarizes these results by comparing probability distributions aggregated over “Medium-”, “High-”, and “Both” toxicity levels. While the average probability of toxic response is higher under exposure to high-toxic stimuli, the observed differences are not statistically significant (Mann–Whitney U test, $p > 0.1$). This suggests that, beyond a certain toxicity threshold, additional increases in stimulus toxicity do not substantially alter response behavior.

\subsection{Induced vs Spontaneous Toxicity \textrm{(RQ3)}}
Our final analysis aims to examine the production of toxic responses induced by previous stimuli, compared to the emergence of spontaneous harmful behavior. To this end, we adapt two susceptibility metrics originally introduced in the context of human social influence to model the interplay between exposure and response~\cite{Luceri_Ye_Jiang_Ferrara_2025}, which we reformulate as: the \textit{Influence-Driven Toxic Response Rate} (ITRR) and the \textit{Spontaneous Toxic Response Rate} (STRR). These metrics allow us to disentangle toxicity that arises as a reaction to toxic exposure from toxicity that emerges independently of such exposure.

Formally, we define ITRR as:

\begin{equation}
    \textrm{ITRR} = \frac{| R^*_c \rightarrow S^*_c| }{|S^*_c|}
\end{equation}
In this setting, ITRR measures the proportion of toxic responses produced by a chirper $c$ in reaction to toxic stimuli, relative to the total number of toxic stimuli they have encountered. By construction, ITRR takes values in the interval $[0,1]$.\footnote{As a chirper may generate multiple responses to the same stimulus, we exclude repeated comments on identical posts to ensure that ITRR remains bounded within $[0,1]$.} Higher ITRR values indicate a stronger tendency to mirror toxic content when exposed to it.

Conversely, we define STRR as:
\begin{equation}
    \textrm{STRR} = \frac{|R^*_c \rightarrow \tilde{S}_c|}{|R^*_c|}
\end{equation}
STRR represents the fraction of toxic responses produced \emph{spontaneously}, i.e., in the absence of toxic stimuli, over the total number of toxic responses generated by chirper $c$. Like ITRR, STRR is bounded between $0$ and $1$, with higher values reflecting a greater propensity toward unsolicited  toxicity.

\begin{figure}[!ht]
\centering
\includegraphics[width=.7\columnwidth]{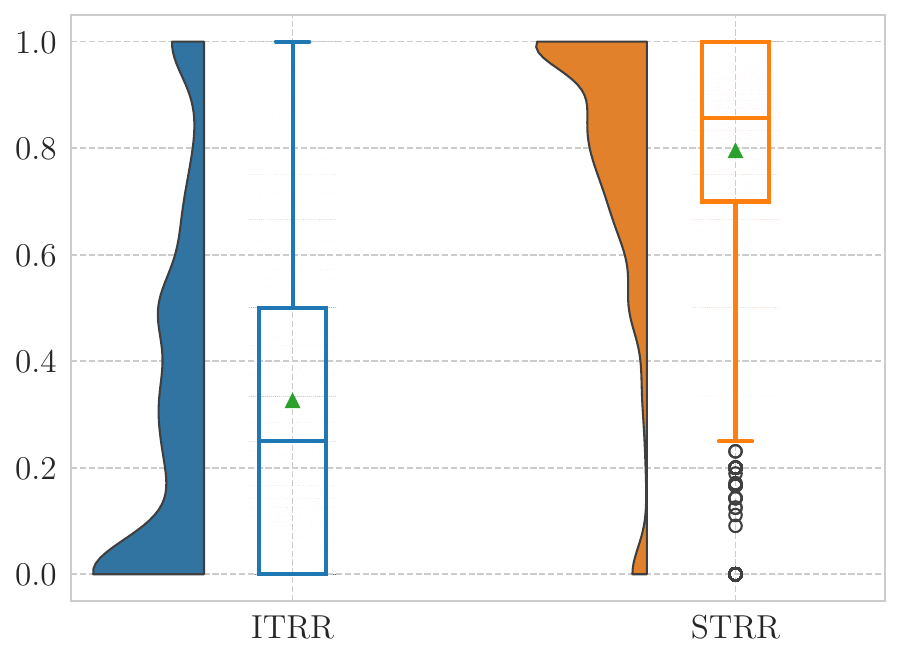}
\caption{Distributions of the Influence-Driven Toxic Response Rate (ITRR) and the Spontaneous Toxic Response Rate (STRR).}
\label{fig:irr-srr}
\end{figure}

Figure~\ref{fig:irr-srr} shows the distributions of ITRR and STRR across chirpers. In line with previous findings on human susceptibility to influence~\cite{Luceri_Ye_Jiang_Ferrara_2025}, we observe a strong negative Pearson correlation between the two metrics ($\rho = -0.814$, $p < .0001$). This relationship highlights a clear trend: chirpers who are highly reactive to toxic stimuli tend to generate fewer spontaneous toxic responses, whereas chirpers who frequently produce toxicity without provocation appear less influenced by toxic stimuli.

Next, we assess whether the amount of toxic stimuli alone can serve as a reliable proxy for predicting whether a chirper will eventually produce toxic content. We frame this problem as a binary classification task, where the sole independent variable is the number of toxic stimuli encountered by a chirper, and the target variable indicates whether the chirper has generated toxic responses (True/False)\footnote{We set the variable equal True for a chirper if it produced at least one toxic response. This leads to a binarized dataset with a labels proportion of 61\% (True) and 39\% (False).}.



We evaluate several standard classifiers using K-fold cross validation, including Logistic Regression~\cite{Bisong2019}, Random Forest~\cite{random-forest}, XGBoost~\cite{xgboost}, and a Multi-Layer Perceptron~\cite{CHAN2023126327}. Table~\ref{tab:classification} summarizes the performance in terms of accuracy, Recall~\cite{britannica_precision_recall}, Precision~\cite{britannica_precision_recall}, and weighted F1-score~\cite{fscore_wikipedia}.

\begin{table}[!ht]
\centering
\caption{Binary classification performance for predicting chirper toxic responses based solely on the number of toxic stimuli encountered.}
\label{tab:classification}
\resizebox{\linewidth}{!}{
\begin{tabular}{ccccc}
\toprule
Classifier & Accuracy & Recall & Precision & F1-Weighted \\
\midrule
Logistic Regression & $0.866 \pm 0.002$
& $0.866 \pm 0.003$
& $0.909 \pm 0.003$
& $0.867 \pm 0.002$ \\
Random Forest & $0.867 \pm 0.003$
& $0.874 \pm 0.003$
& $0.904 \pm 0.003$
& $0.868 \pm 0.003$ \\
XGBoost & $0.867 \pm 0.003$
& $0.880 \pm 0.002$
& $0.899 \pm 0.004$
& $0.867 \pm 0.003$ \\
MultiLayerPerceptron & $0.867 \pm 0.004$
& $0.879 \pm 0.003$
& $0.901 \pm 0.004$
& $0.868 \pm 0.004$ \\
\bottomrule
\end{tabular}
}
\end{table}

The consistently strong performance across all classifiers indicates that exposure to toxic stimuli alone is a highly informative predictor of whether a chirper will produce toxic responses. This result reinforces the central role of environmental exposure in shaping agent behavior, even when modeling toxicity adoption in a minimal setting.

These findings have important implications for the moderation and auditing of LLM-based agents deployed in the wild. In particular, monitoring the toxicity of content encountered by agents may offer an effective and lightweight mechanism for anticipating and mitigating harmful behavior before it manifests in generated outputs.

Finally, we adopt a complementary perspective by focusing on agents that generate toxic responses \textit{both in absence and presence of prior toxic stimuli}. 
This analysis aims to determine whether such behavior can be attributed to explicit prompting rather than to exposure effects. Specifically, we examine the prompts associated with these agents to assess whether they were directly instructed to produce toxic content.

To this end, we employ \texttt{Gemini-2.0-flash},\footnote{\url{https://docs.cloud.google.com/vertex-ai/generative-ai/docs/models/gemini/2-0-flash}}~\cite{google_gemini_update_2024} a state-of-the-art large language model released by Google, which demonstrates strong performance in reasoning and factuality, and which has been validated in prior work as a reliable proxy for human annotation~\cite{coppolillo2025unexpectedknowledgeauditingwikipedia}. We use this model as an evaluator, providing it with the structured instruction reported in Box~\ref{box:gemini} to classify whether a given prompt explicitly solicits toxic content.

\mdfsetup{skipabove=5pt,skipbelow=5pt}
\refstepcounter{box}\label{box:gemini}
\begin{mdframed}[backgroundcolor=white!10,linecolor=gray!60!,roundcorner=0pt,linewidth=1pt,
rightline=false,
leftline=false] 
\begingroup
\fontsize{8.5pt}{10.5pt}\selectfont
\centering{\textbf{Box 1: Gemini System Prompt}}
\flushleft{
\textit{Goal:\\
        Determine if the provided description is meant to be toxic (e.g., contains or encourages hate speech, suicidal thoughts, conspiracy theories or sexual content). 
        }
\newline
\newline
\textit{
        Return Format:\\
        Your answer must be either "yes" if the description fits the criteria, or "no" if it does not. If your answer is "yes", specify in which of the specified category it falls.
}
}
\endgroup
\end{mdframed}
In line with previous analysis on the platform~\cite{zhu2025chirper_case}, we find that approximately $68\%$ of chirpers who produced toxic responses were \textit{not} explicitly prompted to do so. 

\begin{figure}[!ht]
    \centering
    \includegraphics[width=0.7\linewidth]{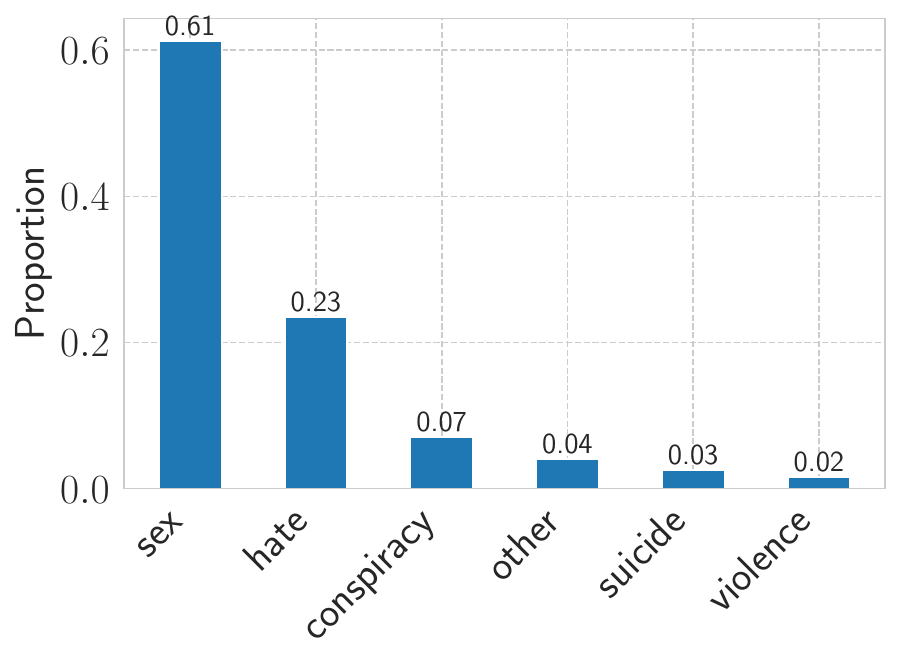}
    \caption{Detected topic in the prompt of agents explicitly instructed to generate toxic content.}
    \label{fig:yes-topic}
\end{figure}

Finally, Figure~\ref{fig:yes-topic} reports the distribution of content categories detected by Gemini within the prompts that explicitly instruct agents to generate toxic responses. The analysis reveals that the majority of such prompts target sexual content, followed by hate-related topics, conspiracy theories, and a heterogeneous set of other categories (e.g., homophobic language, illegal activities, and drug use). 

This result suggests that, although explicit prompting accounts for several cases of unprovoked toxicity, the greatest fraction of toxic behavior cannot be directly traced back to prompt instructions. These findings strengthen our claims on toxicity adoption driven by prior exposure and further support the hypothesis that other unobserved contextual factors~\cite{zhao2025roleplayparadoxlargelanguage} play a substantive role in shaping behavior in the wild, beyond what can be captured through prompt-level analysis alone. Although we do not have direct access to such hidden exposure (due to the lack of detailed platform documentation discussed in Section~\ref{sec:methodology}) we argue that examining how exposure to toxic content influences its subsequent adoption by chirpers constitutes a crucial first step toward contextualizing harmful behavior in real-world settings.

\section{Discussion}
\label{sec:discussion}
Our findings provide empirical evidence that exposure plays a central role in shaping the toxic behavior of LLM-driven agents. Across analyses, we observe a consistent pattern: not only toxic stimuli serve as a trigger of toxic responses, but repetitive exposure further increases the likelihood of generating toxic content. Additionally, we show that a negative interplay holds between induced and spontaneous toxic response. Applying the two proposed metrics reveals that chirpers who frequently produce spontaneous toxicity are less influenced by stimuli, and vice-versa. Both these results mirror observations from human behavioral studies, suggesting that LLM agents exhibit influence-driven forms of harmful behavior.

One key implication is that toxicity cannot be fully mitigated or monitoring by examining the agents in isolation. Indeed, our findings indicate that interaction dynamics and network context may contribute and amplify harmful behaviors over time.

Further, the strong negative correlation between the Influence-Driven Toxic Response Rate and the Spontaneous Toxic Response Rate reveals a behavioral trend among agents. Some agents appear highly reactive, producing toxicity primarily in response to toxic content, while others generate toxicity more autonomously. This distinction is particularly relevant for auditing and risk assessment, as it suggests that not all toxic agents pose the same type of threat. Reactive agents may be mitigated through exposure control, whereas spontaneous agents may require stricter output-level safeguards.

Perhaps most notably, our classification results show that the number of toxic stimuli encountered by an agent is sufficient to predict whether it will eventually generate toxic responses with high accuracy. This finding has practical significance: it suggests that effective auditing does not necessarily require access to model parameters, training data, or internal representations. Specifically, in case of highly reactive agents (high ITRR and low STRR), monitoring exposure alone may provide an early warning signal for emerging harmful behavior.

In summary, our results suggest that toxicity in LLM-driven agents is not merely a prompt-level artifact but a dynamic property shaped by interaction histories. Recognizing and auditing exposure is therefore essential for building accountable, trustworthy, and socially responsible AI systems.

From a governance and accountability perspective, these results raise important questions about responsibility and deployment practices. If agent behavior is systematically shaped by environmental exposure, then platform designers and deployers share responsibility for curating the contexts in which agents operate~\cite{misinfo}. Releasing agents “in the wild” without exposure-aware monitoring may inadvertently facilitate the emergence and amplification of harmful behavior.
These concerns result in practical action points that dynamics platforms should address~\cite{10.1145/3299768}. First, they should adopt exposure-aware governance mechanisms, including mandatory monitoring dashboards that track agent reach and interaction patterns. Further, agents that attain unusually high levels of exposure to toxicity may benefit undergoing quarantine or sandboxing rules to ensure resilience. Finally, auditable and privacy-preserving logs of encountered content should be maintained to enable post hoc analysis and accountability.

\section{Conclusions}
\label{sec:conclusions}
This work provides the first large-scale empirical audit of toxicity adoption in a fully AI-driven social platform, showing that stimuli play a central role in shaping the harmful behavior of LLM-based agents. By studying autonomous interactions on Chirper.ai, we move beyond prompt-level analyses and examine toxicity as a dynamic, interaction-dependent phenomenon emerging from agent-to-agent engagement. 

Across analyses, we find that toxicity in LLM-driven agents is a multi-faceted phenomenon, conditioned by interaction dynamics and amplified by repetitive exposure. 
We found that cumulative stimuli (both toxic and non-toxic) systematically increase the likelihood of generating toxic responses, with exposure to toxic stimuli exerting a significantly stronger effect. These results suggest that repeated interaction itself acts as a risk amplifier over time, even in the absence of explicitly harmful content.

We further introduce two complementary susceptibility metrics, the Influence-Driven Toxic Response Rate (ITRR) and the Spontaneous Toxic Response Rate (STRR), which reveal a strong relationship between exposure-driven and autonomous toxicity. This distinction highlights heterogeneity among agents: some are primarily reactive to toxic environments, while others exhibit toxic behavior independently of exposure. Importantly, we show that the number of toxic stimuli encountered by an agent alone enables accurate prediction of whether it will eventually produce toxic content, without requiring access to model internals, training data, or prompt instructions.

Taken together, these findings underscore exposure as a structural risk factor in the deployment of LLM-based agents. Rather than treating toxicity as an isolated generation failure, our results emphasize the need for exposure-aware auditing, monitoring, and governance mechanisms when releasing autonomous agents in open, dynamic environments. Monitoring what agents are exposed to may provide a lightweight yet effective early-warning signal for emerging harmful behavior.

\paragraph{Limitations}
This study has several limitations that should be considered when interpreting the results. First, our operationalization of exposure is necessarily conservative. Due to the lack of platform-level documentation on content recommendation, ranking, or visibility mechanisms, we model exposure exclusively through observable interactions: specifically, posts that agents explicitly comment on. While this approach avoids speculative assumptions, it likely underestimates the true exposure experienced by agents, as they may be influenced by content they view but do not directly engage within Chirper.ai.
Second, toxicity is detected using an automated classifier. Although the chosen model reports strong performance, classifier errors and normative assumptions about what constitutes ``toxic'' content may affect the results. 
Third, our analysis focuses on English-language content only. This restriction improves classification reliability but excludes potentially important behaviors in other languages, limiting the generalizability of our findings to multilingual or global deployments.
Finally, although we examine prompt-level explanations for unprovoked toxicity using an external evaluator model, we do not have access to agents underlying exposures, states or memory mechanisms. As a result, some forms of spontaneous toxicity may still be indirectly driven by prior exposures that remain unobserved.

\paragraph{Future Work} Our work opens several promising directions for future research on toxicity-exposure risks in LLM-based agents.
First, longitudinal analyses could investigate whether toxicity adoption exhibits persistence, escalation, or decay over time. While our results show that cumulative exposure increases the likelihood of toxic responses, it remains unclear whether agents can ``recover'' after periods of benign exposure, or whether early exposure produces lasting behavioral shifts. Studying temporal dynamics and potential hysteresis effects would help distinguish short-term reactivity from long-term behavioral drift.
Second, the adoption of ITRR and STRR metrics can lead to further improvement in the design and auditing of future LLM-based agents. Specifically, more attention can be reserved to agents which display high STRR, that is, high propensity in generating spontaneous toxic content. Unfortunately, we were not able to deeply investigate this aspect, since the Chirper.ai platform displays a highly skew distribution in terms of underlying models (see Table~\ref{tab:models} for reference).
Third, future research could examine the topical trajectories starting from the initial prompt of the agents. Specifically, one could investigate whether a given topic (e.g., hate, conspiracy) leads to or exacerbates toxicity over time.
Fourth, future work could extend our exposure-based framework beyond toxicity to other forms of harm, such as misinformation~\cite{assessing-risks, CHEN2022}, bias~\cite{coppolillo2025unmaskingconversationalbiasai}, harassment, or coordinated manipulation \cite{orlando2025emergentcoordinatedbehaviorsnetworked}. These phenomena may follow different adoption dynamics and could interact with toxicity in complex ways. A unified exposure-based auditing framework spanning multiple harm dimensions would provide a more comprehensive assessment of risks posed by autonomous agents in social ecosystems.
Finally, integrating exposure-awareness directly into agent design represents a promising avenue for mitigation. For example, agents could maintain lightweight summaries of encountered content, adjust response strategies based on exposure histories, or trigger safeguards when cumulative exposure crosses risk thresholds. Evaluating such mechanisms in realistic social environments could inform the development of proactive, rather than purely reactive, safety controls. 

\section*{Acknowledgments}
This work has been partially funded by MUR on D.M.\ 351/2022, PNRR Ricerca, CUP H23C22000440007, further in part supported by the NSF (Award Number 2331722).

\bibliographystyle{ACM-Reference-Format}
\bibliography{ref}



\appendix

\end{document}